\documentclass[preprint,pre]{revtex4}
\usepackage{amssymb}
\usepackage{amsmath}
\usepackage{epsfig}

\def\ket#1{\mathinner{|{#1}\rangle}}

\def\expect#1{\langle#1\rangle}

\def\id3{ {\mathbf{1}_{3\times 3}}}
\def\ln {\text{ln}}
\def\e {\text{e}}
\def\D {\text{d}}
\begin{document}

\title{Phase transitions and correlations in the bosonic pair contact process with diffusion:
Exact results\\
}
\author{Matthias Paessens}
\email{m.paessens@fz-juelich.de}
\author{Gunter M. Sch\"utz}

\affiliation{Institut f\"ur Festk\"orperforschung, Forschungszentrum J\"ulich - 
52425 J\"ulich, Germany}
\date{\today}

\begin{abstract}
{
The variance of the local density of the pair contact process with diffusion (PCPD) 
is investigated
in a bosonic description. At the critical point of the absorbing phase transition (where
the average particle number remains constant) it is shown that for lattice dimension $d>2$
the variance exhibits a phase transition: For high enough diffusion constants,
it asymptotically approaches a finite value, while for low diffusion
constants the variance diverges exponentially in time. This behavior appears also
in the density correlation function, implying that the 
correlation time is negative. Yet one has 
dynamical scaling with a dynamical 
exponent calculated to be $z=2$.
}

\end{abstract}

\maketitle

\section{Introduction}
An prototypical example for critical phenomena in nonequilibrium statistical physics is the
absorbing phase transition. This is a transition from an active fluctuating
phase with a finite particle density to an absorbing state where any dynamics
is suppressed. One has found rather robust universality classes, e.g. the class
of directed percolation (DP) and the parity conserving universality class (PC).
A member of the DP--class is the pair contact process where two
neighboring particles may create an offspring on a third lattice site or 
may annihilate each other. 

This model extended by particle diffusion -- the pair contact process
with diffusion (PCPD) -- has attracted much interest because it is not known
to which universality class it belongs. Several possibilities have been discussed:
It was found that some exponents are very close to those of the PC class
\cite{car00}, more recent investigations however give hints for a DP
behavior \cite{bar00}. It was also suggested that the 
critical behavior of the PCPD defines a new universality class \cite{odo01,koc00},
or may depend on the diffusion constant \cite{odo00}.  
Analytical results are rare 
in this field and one has to revert to numerical methods like the density
matrix renormalization group (DMRG) or Monte Carlo simulations.

This is different for the bosonic description of the model where the
exclusion interaction -- which constraints the number of particles at one
site to at most one -- is dropped. In this case a field theoretic approach
due to Howard and T\"auber \cite{how00} is available. A drawback of this
approach is that it is not suitable for deciding the universality
class of the model with particle number restriction. 
In this paper we show by an exact treatment of the model 
that the diffusion constant and the lattice dimension have considerable
impact on the phase transition and correlations 
of the bosonic PCPD. Although the particle exclusion
interaction is crucial for the behavior of the system
this investigation gives some insight on the role
of diffusion in the PCPD.

\section{Model}
We define the following process:
On a infinite $d$--dimensional cubic lattice particles ('$A$') are diffusing with rate $D$,
 in each spatial direction. Additionally
they branch and annihilate:
$k\ge 1$ particles $A$ are created with rate $\mu$ out of any set of $m\ge 1$ particles 
($m$ fixed), and
$l\ge 1$ particles are annihilated with rate $\lambda$ out of any set of $p\ge l$ particles
($l$ fixed):
\begin{eqnarray}
mA &\overset{\mu}{\rightarrow}&(m+k)A \nonumber \\
pA &\overset{\lambda}{\rightarrow}&(p-l)A \nonumber \\
A\cdot & \overset{D}{\leftrightarrow}&\cdot A.
\end{eqnarray}
The number of particles on each lattice site is not restricted -- the creation and 
annihilation processes take place on one lattice site. Thus the bosonic representation
of the process is used.  We try
to keep the description as general as possible, but as we will see, analytical results
are available only for few cases. In this paper we investigate the two cases where
$p=m=1$ or $p=m=2$ and arbitrary $k$ and $l\le p$.
One special case is the PCPD, where $m=p=l=2$ and $k=1$.

Following the notation and formalism introduced in \cite{doi00,tri00} we define the site
occupation numbers as $\vec{n}=\left\{n({\bf x})\right\}$. Then the time dependent probability
vector describing the system can be expressed as
\begin{equation}
\ket{F(t)}=\sum_{n({\bf x})} P(\vec{n},t) \ket{\vec{n}}
\end{equation}
where the $\ket{\vec{n}}$ are the basis vectors spanning the state space and $P(\vec{n},t)$
is the probability distribution of the site occupation numbers. 
The master equation describing the time evolution of the probability distribution can then
be written as
\begin{equation}
\frac{\partial}{\partial t} \ket{F(t)} = - {\cal H} \ket{F(t)},
\end{equation}
${\cal H}$ is the stochastic generator of the system, often called as ``hamiltonian'' due
to the analogy of the master equation to the Schr\"odinger equation (in imaginary time)
\cite{sch00}.
Let $a({\bf x})$ and $a({\bf x})^\dagger$ be the space dependent annihilation and creation 
operators and $n({\bf x})=a^\dagger({\bf x})a({\bf x})$ the particle number operator, then
the hamiltonian is given by
\begin{eqnarray}
{\cal H}& =& - D \sum_{k=1}^{d} \sum_{{\bf x}} \left[
a({\bf x})a^\dagger({\bf x}+{\bf k}) + a^\dagger({\bf x})a({\bf x}+{\bf k})
                                      -2 n({\bf x})\right]\nonumber \\
&& - \lambda \sum_{{\bf x}} \left[\left(a^\dagger({\bf x})\right)^{(p-l)} \left(a({\bf x})\right)^p
                                 -   \prod_{i=1}^{p} \left( n({\bf x}) -i +1\right) 
                           \right] \nonumber \\
&& - \mu \sum_{{\bf x}} \left[\left(a^\dagger({\bf x})\right)^{(m+k)} \left(a({\bf x})\right)^m
                                 - \prod_{i=1}^{m} \left( n({\bf x}) -i +1\right) 
                           \right],
\end{eqnarray}
where ${\bf k}\equiv {\bf k}(k)=(\ldots,0,1,0,\ldots)^T$ is the $k$-th unit space vector.
The time evolution of an operator $b({\bf y})$ is calculated by
\begin{equation}
\frac{\partial}{\partial t} {b({\bf y})} = {\left[{\cal H},b({\bf y})\right]}.
\end{equation}
Using the commutator rule $\left[ a({\bf x}), a^\dagger({\bf y}) \right]=\delta_{\bf x,y}$
we get after straightforward calculations  
\begin{eqnarray}
\frac{\partial}{\partial t} \expect{a({\bf x})} &=& D \sum_{k=1}^{d} 
      \left\{     \expect{a({\bf x}-{\bf k})} + \expect{a({\bf x}+{\bf k})}
                -2\expect{a({\bf x})} 
      \right\}  \label{eq.4a} \\ 
&& - \lambda l \expect{a({\bf x})^p} + \mu k \expect{a(\bf x)^m}\nonumber \\
\frac{\partial}{\partial t} \expect{a({\bf x})a({\bf y})} 
                                         & \underset{{\bf x}\ne {\bf y}}{ = }&
  D \sum_{k=1}^{d} \left\{\right. \expect{ a({\bf x})a({\bf y}-{\bf k}) } + 
                           \expect{ a({\bf x})a({\bf y}+{\bf k}) } + \label{eq.4b} \\
                 &&         \expect{ a({\bf x}-{\bf k})a({\bf y}) } +
                           \expect{ a({\bf x}+{\bf k})a({\bf y}) } -
                         4 \expect{ a({\bf x})a({\bf y})} \left.\right\} \nonumber\\
  && - \lambda l   \left\{\right. \expect{ a({\bf x})   a({\bf y})^p } + 
                                 \expect{ a({\bf x})^p a({\bf y})   } \left.\right\} \nonumber\\
&&  + \mu k        \left\{\right. \expect{ a({\bf x})   a({\bf y})^m } + 
                                 \expect{ a({\bf x})^m a({\bf y})   } \left.\right\}\nonumber\\
\frac{\partial}{\partial t} \expect{\left(a({\bf x})\right)^2} &=& 
     2D \sum_{k=1}^{d} \left\{ \expect{ a({\bf x})a({\bf x}-{\bf k}) } +
                               \expect{ a({\bf x})a({\bf x}+{\bf k}) } -
                              2\expect{ a({\bf x})^2} \right\} \label{eq.4c}\\
&& + \lambda l \left\{\right.\,\, (1+l-2p) \,\,\expect{a({\bf x})^p} 
                                  - 2 \expect{ a({\bf x})^{p+1}} \left.\right\}\nonumber\\
&& - \mu k     \left\{\right. (1-k-2m) \expect{a({\bf x})^m}
                                  - 2 \expect{ a({\bf x})^{m+1}} 
                     \left.\right\} \nonumber
\end{eqnarray}
 
Using $\expect{n({\bf x})}=\expect{a({\bf x})}$ and 
$\expect{n({\bf x})^2}=\expect{a({\bf x})^2}+\expect{a({\bf x})}$ this set of coupled
difference--differential equation allows for the analytical calculation of the time--dependent
expectation value of the particle density and its autocorrelation in some special cases.

We restrict to the case $p=m$ where the creation and annihilation processes are balanced 
and an absorbing phase transition can be found. 
For $\lambda l > \mu k$ the particles die out exponentially ($p=m=1$) or according
to a power law ($p=m>1$), while for $\lambda l < \mu k$ the particle density 
diverges. 
Here a crucial difference between the description with and without particle
number restriction can be seen: While in the models with exclusion interaction the
absorbing phase transition is of second order, the bosonic model exhibits a 
first order transition. 

In analogy to the exclusion model we call the
rate which divides the two different behaviors the ``critical'' rate, which
from Eq.~(\ref{eq.4a}) can be read off as
\begin{equation}
\lambda_c=\mu k /l
\end{equation}
for given $\mu$. 
For this rate the particle density 
is constant for all times $\expect{a({\bf x},t)}=\rho_0$ (for homogeneous initial 
conditions), as can be seen from
Eq.~(\ref{eq.4a}) which reduces to a diffusion equation.
Thus the interesting quantity
is the variance $\sigma^2=\expect{n({\bf x})^2} - \expect{n({\bf x})}^2$
 which we shall investigate in what follows. 

Eliminating $p$ and $\lambda$ in Eqs.~(\ref{eq.4a})-(\ref{eq.4c}) one gets
\begin{equation}\begin{split}
\frac{\partial}{\partial t} \expect{a({\bf x})} =& D \sum_{k=1}^{d} 
      \left\{     \expect{a({\bf x}-{\bf k})} + \expect{a({\bf x}+{\bf k})}
                -2\expect{a({\bf x})} 
      \right\}  \\
\frac{\partial}{\partial t} \expect{a({\bf x})a({\bf y})} 
                                          \underset{{\bf x}\ne {\bf y}}{=} &
  D \sum_{k=1}^{d} \left\{\right. \expect{ a({\bf x})a({\bf y}-{\bf k}) } + 
                           \expect{ a({\bf x})a({\bf y}+{\bf k}) } +\\
                 &         \expect{ a({\bf x}-{\bf k})a({\bf y}) } +
                           \expect{ a({\bf x}+{\bf k})a({\bf y}) } -
                         4 \expect{ a({\bf x})a({\bf y})} \left.\right\} \\
\frac{\partial}{\partial t} \expect{\left(a({\bf x})\right)^2} =& 
     2D \sum_{k=1}^{d} \left\{ \expect{ a({\bf x})a({\bf x}-{\bf k}) } +
                               \expect{ a({\bf x})a({\bf x}+{\bf k}) } -
                              2\expect{ a({\bf x})^2} \right\} \\
& +\mu k (k+l) \expect{a({\bf x})^m} 
\label{eq.5}
\end{split}\end{equation}

We see that this set of equations is only closed for the cases $m=1$ or $m=2$.

In the case of a vanishing diffusion constant, $D=0$, the lattice sites
are independent of each other. Thus the description of the process reduces 
to the zero dimensional case $d=0$,
\begin{equation}\begin{split}
\frac{\partial}{\partial t} \expect{a({\bf x})} =& 0 \\
\frac{\partial}{\partial t} \expect{\left(a({\bf x})\right)^2} =& \mu k (k+l)
\expect{a({\bf x})^m} ,
\label{eq.5.5}
\end{split}\end{equation}
and has to be treated separately.

\subsection{Contact process with diffusion, $m=1$}

Here, only $l=1$ is possible. Additionally by rescaling $\mu$ we may fix $k=1$.
This case has already been considered in \cite{hou00} as a model for clustering
of biological organisms \cite{you00}. For convenience we summarize
the main results here.
 
For $D=0$ or $d=0$ Eq.~(\ref{eq.5.5}) directly yields $\expect{a({\bf x})^2}=c_0+c_1\,\, t$ and
thus the variance diverges. 
For $D\ne 0$ the fluctuations of the particle density diverges for dimensions $d\le 2$ while
they remain finite for $d>2$,

\begin{equation}
\expect{a({\bf x})^2}=\left\{\begin{array}{cl}
      c_1 \,\, t^{-d/2+1} & d<2 \\
      c_2 \,\, \ln t       & d=2 \\
      c_3 + c_4 \,\, t^{-d/2+1} & d>2 
    \end{array} \right.
\end{equation}
where $t\gg 1$ and $c_0, \ldots, c_4$ are positive constants.

\subsection{Pair contact process with diffusion, $m=2$}

We now derive analytically the late time behavior of the solution for $m=2$.

For $D=0$ or $d=0$ Eq.~(\ref{eq.5.5}) yields 
\begin{equation}
\begin{split}
\expect{a({\bf x}^2)}=\rho_0^2 \exp(t/\tau), \\
\tau=\frac{1}{\mu k (k+l)}.
\label{eq.A}
\end{split}
\end{equation}
The variance diverges exponentially in time as opposed to $m=1$ where the divergence is
linear. Only for times small compared to $\tau$ the variance Eq.~(\ref{eq.A}) grows linearly.

For $D\ne 0$ 
we get the solution by applying Fourier-- and Laplace--transformations. 
We also present the crossover from short to late time behavior, which has to be calculated
numerically.

First we rescale time by
\begin{equation}
t\to \frac{t}{2D},
\end{equation}
and define
\begin{equation}
\begin{split}
F_{\bf x}({\bf r},t)=&\expect{a({\bf x})a({\bf x+r})} = \expect{ n({\bf x}) n({\bf x+r}) } 
                                              -\delta_{\bf r,0} \expect{n({\bf x})}\\
\alpha=&\frac{\mu k (k+l)}{2D}.
\end{split}
\end{equation}

The parameter $\alpha$ is a measure for the weighting of reaction rates to diffusion, small
$\alpha$ corresponds to dominant diffusion, while large $\alpha$ corresponds to
dominating reaction rates.
In what follows we consider only translational invariant inital conditions, in which case
 $F_{\bf x}({\bf r},t)$ is independent of ${\bf x}$. 
Using Eq.~(\ref{eq.5}) we get the following difference--differential equation for $F$: 
\begin{equation}
\begin{split}
\frac{\partial}{\partial t} F({\bf r},t)=&\sum_{k=1}^{d}
          \left\{F({\bf r-k},t)+F({\bf r+k},t)-2F({\bf r},t)\right\}  +
              \delta_{\bf r,0} \alpha F({\bf 0},t)\\
=&\sum_{k=1}^{d}
          \Delta_k F({\bf r},t) +
              \delta_{\bf r,0} \alpha F({\bf 0},t)
\end{split}
\label{eq.9}
\end{equation}
where $\Delta_k$ is the discrete Laplacian concerning
the $k$--th component. The variance $\sigma^2$ is related to $F$ as follows
\begin{equation}
\sigma(t)^2=F({\bf 0},t)+\rho_0-\rho_0^2.
\end{equation}

Here, we see that there is no qualitative difference between parity conserving models 
($k$ and $l$ even) 
and non--parity conserving models --- models with different $k$ and $l$ 
differ only by different creation and annihilation rates.

This kind of equation can be solved using the Fourier--transformation:
\begin{equation}
f({\bf q},t)=\sum_{\bf r} \e^{-i{\bf qr}} F({\bf r},t), 
\qquad F({\bf r},t)=\int\,\frac{\D^d{\bf q}}{(2\pi)^d}\e^{i{\bf qr}} f({\bf q},t).
\end{equation}
We get
\begin{equation}
\frac{\partial}{\partial t} f({\bf q},t)= - w({\bf q}) f({\bf q},t) + \alpha F({\bf 0},t),
\end{equation}
with the dispersion relation $w({\bf q})= -2 \sum_{k=1}^{d}\left( \text{cos}\left(q_k\right)-1 \right)$.
Integration yields
\begin{equation}
f({\bf q},t)=\e^{-w({\bf q})t}\, \left\{ f({\bf q},0) + 
     \alpha \int_{0}^{t}\!\D\tau F({\bf 0},\tau) \e^{w({\bf q})\tau} \right\}.
\end{equation}
As initial condition we choose a Poisson--distribution $F({\bf r},0)=\rho_0^2$ so that
$f({\bf q},0)=\delta_{\bf q,0} \rho_0^2$. Thus we get
\begin{equation}
F({\bf r},t)=\rho_0^2 + \alpha \int_{0}^{t}\!\D\tau F({\bf 0},\tau) b({\bf r},t-\tau)
\label{eq.13}
\end{equation}
with
\begin{equation}
\begin{split}
b({\bf r},t)=& \int\!\frac{\D^d{\bf q}}{(2\pi)^d} \e^{-w({\bf q})t+i{\bf qr}}\\
=& \e^{-2dt} I_{r_1}(2t)\cdot \ldots \cdot I_{r_d}(2t)
\end{split}
\label{eq.14}
\end{equation}
where $I_r(t)$ is the modified Bessel function of order $r$. The dimension $d$ is 
now just a parameter which can formally take real values. Although this is not
physical it allows for the investigation of the dependence on the dimension.

For ${\bf r=0}$ the long--time behavior of the solution of the Volterra integral--equation 
Eq.~(\ref{eq.13}) with the function $b(t)$ given by
Eq.~(\ref{eq.14}) is known from the mean spherical model
\footnote{In the mean spherical model the spherical constraint is parametrized mathematically
by a Lagrangian multiplier. This multiplier is determined by the Volterra integral--equation
Eq.~(\ref{eq.13}) where $\rho_0^2$ is replaced by $b({\bf 0},t)$ which does not change
the long time--time behavior.}. In this context $\alpha$ plays the role of the temperature.
This analogy enables us to use known results from
the spherical model. 
Eq.~(\ref{eq.13}) can be solved using temporal Laplace transformation \cite{god00},
\begin{equation}
\tilde{F}(p)=\int_{0}^{\infty}\!\D t\, \e^{-pt} F({\bf 0},t).
\end{equation}
We get
\begin{eqnarray}
\tilde{F}(p)&=&\frac{\rho_0^2}{p} + \alpha \tilde{F}(p) \tilde{b}(p) \nonumber \\
\Leftrightarrow \tilde{F}(p)&=&\frac{\rho_0^2}{p(1-\alpha \tilde{b}(p))}.
\label{eq.16}
\end{eqnarray}

for late times $F({\bf 0},t)$ is given by the behavior of $\tilde{b}(p)$ for small $p$, which
crucially depends on the dimension $d$ (see for example \cite{god00}):
\begin{equation}
\begin{split}
\tilde{b}(p)=&\left\{\begin{array}{ll}
(4\pi)^{-d/2} \Gamma(1-d/2) p^{-(1-d/2)} & d<2 \\
2 A_1 - (4\pi)^{-d/2} \vert \Gamma(1-d/2)\vert p^{d/2-1} & 2<d<4\\
2 A_1 - 4 A_2 p & d>4 
\end{array} \right.  \\
A_k=& \int\!\frac{\D^d{\bf q}}{(2\pi)^d}\frac{1}{(2w({\bf q}))^k}
\end{split}
\end{equation}
This results in different behavior of $F({\bf 0},t)$ as we shall see in the 
next sections. \\
For all even integral dimensions $d=2,4,\ldots$ logarithmic corrections arise whose investigation
goes beyond the scope of this paper. 

\subsubsection{${\bf d<2}$}

\begin{figure}[l]
\centerline{\epsfxsize=3.5in\epsfbox
{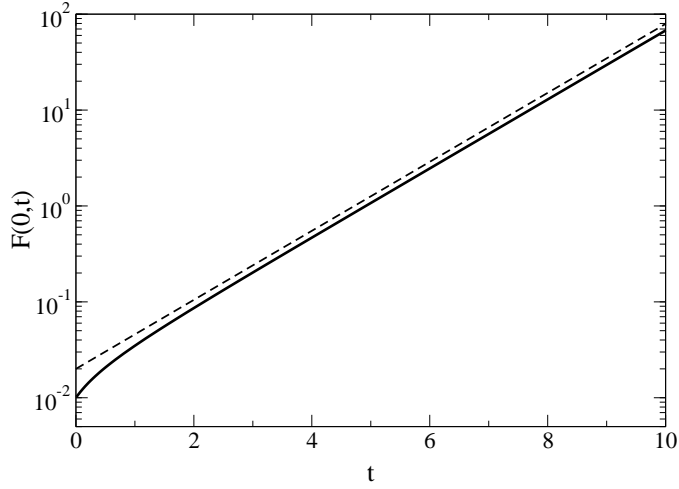}
}
\caption{Numerical calculation of $F({\bf 0},t)$ for $d=1, \alpha=2, \rho_0=0.1$. The dashed
line shows the theoretical predicted slope $\tau\approx 1.2071$.
\label{Fig1}}
\end{figure}
%
%
As for $d<2$ the quantity $\tilde{b}(p)$ diverges for $p\to 0$ the denominator of Eq.~(\ref{eq.16}) 
has always  a zero for $p\ne 0$, so that $\tilde{F}(p)$ has  a pole at a positive value $p=1/\tau$.
A pole of the laplace transform corresponds to exponential behavior of the
original function and we get
\begin{equation}
F({\bf 0},t)\underset{t\to\infty}{\propto} \e^{t/\tau}.
\label{eq.19}
\end{equation}  
For $d=1$ the exact expression of $\tilde{b}$ is known \cite{god00}:
\begin{equation}
\tilde{b}(p)=\frac{1}{\sqrt{p(p+4)}}
\end{equation}
which yields
\begin{equation}
\tau_{d=1}=\frac{1}{\sqrt{4+\alpha^2}-2}.
\end{equation}
For any finite value of $\alpha$ the time scale $\tau$ is finite but diverges
if $\alpha\searrow 0$. This is in analogy to the spherical model, where in 
one dimension the critical temperature is zero. \\

In order to investigate how the predicted asymptotic behavior for large times
is approached we have performed a numerical integration of $F({\bf 0},t)$, shown 
in Fig.~\ref{Fig1}. For details of the numerical calculation see \cite{pae00}, where a similar 
integral equation is calculated. We see that the asymptotic behavior is approached quickly
and the solution Eq.~(\ref{eq.19}) is a good approximation for times $t>1$.

\subsubsection{${\bf 2<d<4}$}

\begin{figure}[l]
\centerline{\epsfxsize=3.5in\epsfbox
{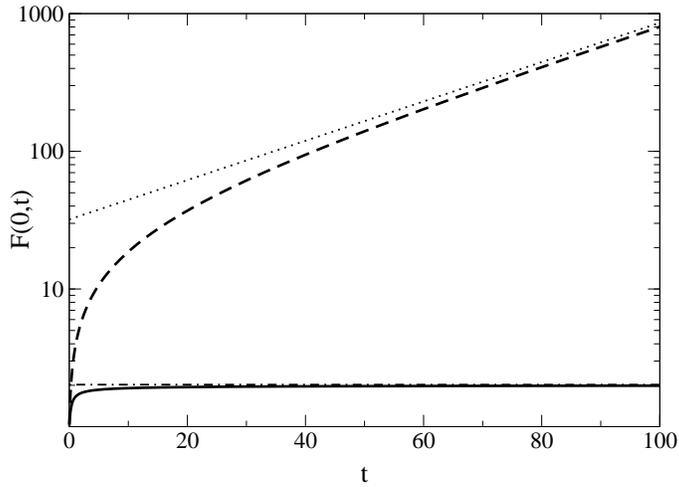}
}
\caption{Numerical calculation of $F({\bf 0},t)$ for $d=3, \rho_0=1$, $\alpha=2 <\alpha_c$ (solid line) and
$\alpha=4.2 > \alpha_c$ (dashed line). The dotted
line shows the theoretical predicted slope $\tau\approx 30.4$, the dashed--dotted line 
the theoretical predicted asymptotic value $F({\bf 0},t=\infty)\approx 2.02$.
\label{Fig2}}
\end{figure}
%
%
For $d>2$ the quantity $\tilde{b}(p)$ shows qualitatively different behavior: It approaches the 
finite value $2A_1$ for $p\to 0$. Therefore the $\tilde{F}(p)$ has a pole 
for positive $p$ 
only for $\alpha$ larger
than a critical value given by 
\begin{equation}
\alpha_c=\frac{1}{2A_1},
\end{equation}
which is identical to the critical temperature in the 
spherical model.
Thus we find a phase transition in the behavior of the autocorrelation 
$F({\bf 0},t)$:
For $\alpha>\alpha_c$ (low diffusion constant) we recover the exponential 
divergence
\begin{equation}
F({\bf 0},t)\underset{t\to\infty}{\propto} \e^{t/\tau}.
\end{equation} 
with a time scale
\begin{equation}
\tau=\left( \frac{ \alpha^\prime}{c_2 \alpha}\right) ^{-\frac{1}{d/2-1}}
\label{eq.24}
\end{equation}
with the reduced control parameter
\begin{equation}
\alpha^\prime=\frac{\alpha-\alpha_c}{\alpha_c},
\end{equation}
and $c_2=(4\pi)^{-d/2} \vert\Gamma(1-d/2)\vert$. 
This time scale diverges if we approach
$\alpha\searrow\alpha_c$.
 
For $\alpha<\alpha_c$ (high diffusion constant) the pole of $\tilde{F}(p)$ 
vanishes and $F({\bf 0},t)$ 
asymptotically approaches a finite value
\begin{equation}
F_{\infty}=\lim_{t\to\infty}F({\bf 0},t) = \frac{\rho_0^2}{1-\alpha/\alpha_c} > \rho_0^2
\label{eq.23}
\end{equation}
which diverges if we approach $\alpha\nearrow\alpha_c$. 

Therefore a suitable order parameter for this phase transition is  
$F_{\infty}^{-1}$ which decreases linearly to zero for 
$\alpha\nearrow\alpha_c$ and is equal to zero for $\alpha>\alpha_c$.

For $\alpha=\alpha_c$ we get
\begin{equation}
\tilde{F}(p)=\frac{(4\pi)^{d/2}\rho_0^2}{\left\vert\Gamma(1-d/2)\right\vert\alpha_c}
   \frac{1}{p^{d/2}}
\end{equation}
which results in a power law 
\begin{equation}
F({\bf 0},t)\propto t^{d/2-1}
\end{equation}
and hence in a power law divergence of the variance.

\subsubsection{${\bf d>4}$}

For $d>4$ we find qualitatively the same behavior as for $2<d<4$.
Like in the previous case, $\tilde{F}(p)$ has a pole at a positive $p$ only for
values $\alpha>\alpha_c=1/(2A_1)$. For $\alpha>\alpha_c$ the time
scale of the exponential increase is given by
\begin{equation}
\tau=\left(\frac{\alpha^\prime}{4A_2 \alpha}\right)^{-1}.
\label{eq.tau_mf}
\end{equation}
The difference to the case $2<d<4$ is, that this time scale is now 
independent of the dimension $d$, indicating that we are in the mean field region.

For $\alpha<\alpha_c$, $F({\bf 0},t)$ approaches the asymptotic value given by
Eq.~(\ref{eq.23}).

For $\alpha=\alpha_c$ we get
\begin{equation}
\tilde{F}(p)=\frac{\rho_0^2}{4A_2 \alpha_c p^2}
\end{equation}
which results in a power law
\begin{equation}
F({\bf 0},t)\propto t.
\end{equation}

\begin{figure}[l]
\centerline{\epsfxsize=2in\epsfbox
{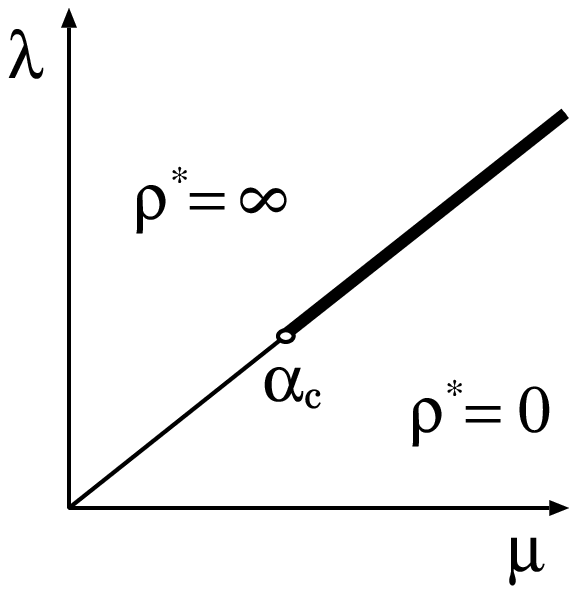}
}
\caption{The phase diagram of the system for fixed diffusion constat $D$: 
In the limit of $t\to\infty$ for $\lambda<\mu k/l$ 
the stationary density $\rho^*$ is zero while it diverges for $\lambda>\mu k/l$. 
For $\lambda=\mu k/l$ the density is constant, $\rho^*=\rho_0$, and the variance
function is bounded for $\alpha<\alpha_c$, while it diverges exponentially for $\alpha>\alpha_c$
and algebraically for $\alpha=\alpha_c$, where $\alpha=\mu k (k+l)/(2D)$.
\label{Fig_pd}}
\end{figure}

These results are summed up in the phase diagram Fig.~\ref{Fig_pd}.
%
%

\section{Spatial correlations}
\begin{figure}[l]
\centerline{\epsfxsize=3.5in\epsfbox
{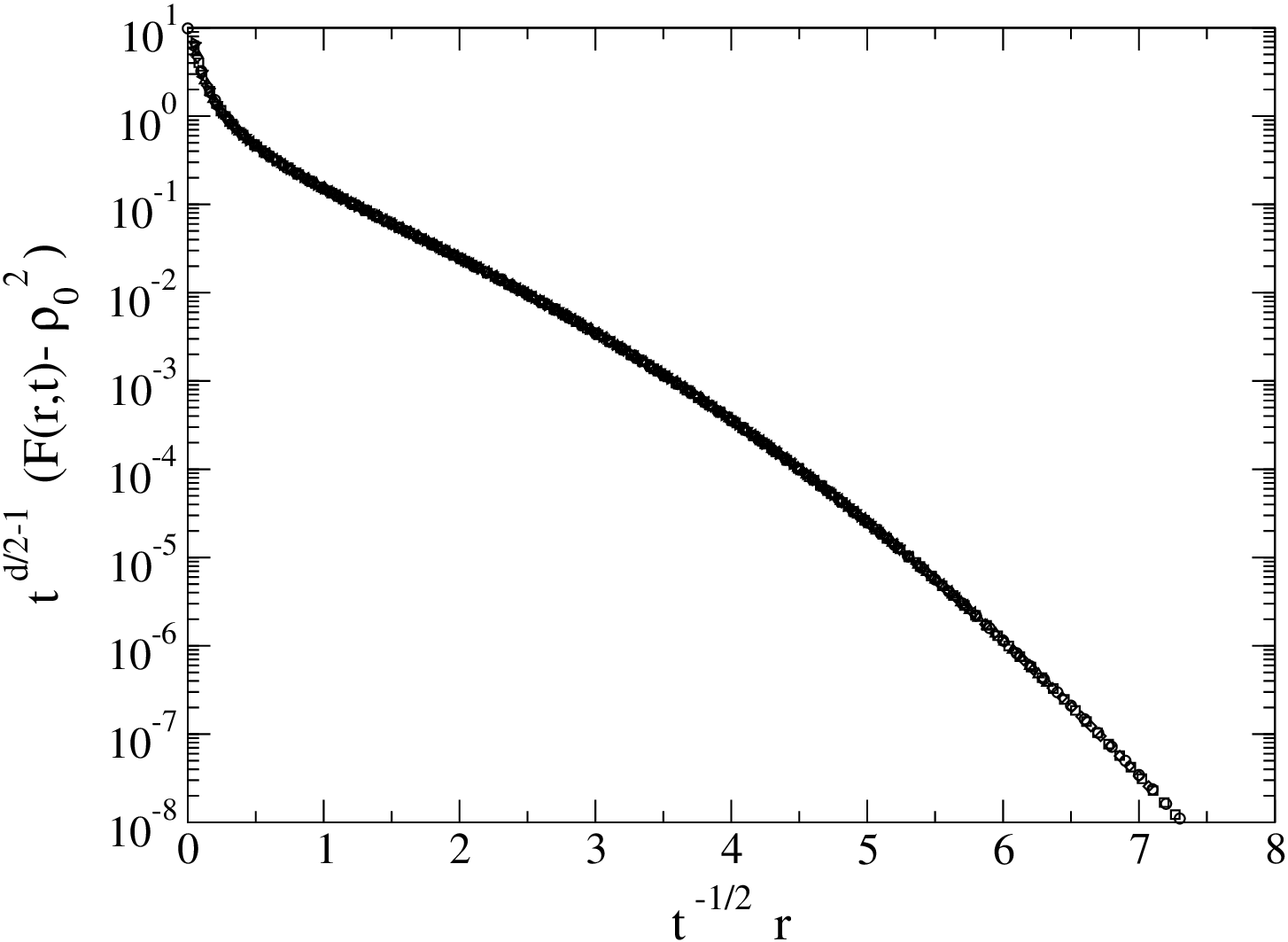}
}
\caption{Numerical calculation of $F({\bf r}=(r,0,...),t)$ for 
$d=3, \alpha=2<\alpha_c, \rho_0=1$ and
times $t=50,100,150,\ldots,400$.
\label{Fig3}}
\end{figure}
%
%
In the mean field regime ($d>4$) the behavior of the correlation function $F({\bf r},t)$
can be calculated analytically in the limit of large ${\bf r}$ and $t$. As derived
in the appendix we get:
\begin{equation}
F({\bf r},t)-\rho_0^2=\left\{\begin{array}{cl}
      \frac{\rho_0^2\alpha}{4\pi^{d/2} \vert \alpha^\prime \vert} \,\,
          r^{2-d}\, \Gamma\left(\frac{d}{2}-1,\frac{r^2}{4t}\right)  & \alpha^\prime<0  \\ 
\\
\frac{\rho_0^2}{64 A_2  \pi^{d/2}}\,\,r^{4-d}\,\, \Psi\left(d,\frac{r^2}{4t}\right) & \alpha^\prime=0\\ 
\\

  \frac{\rho_0^2}{\left(8\pi\right)^{(d-1)/2}A_2^{(d-2)/2}}
     \left(\frac{\alpha^\prime}{\alpha}\right)^{(d-4)/2}
         \,\left(\frac{r}{\xi}\right)^{(1-d)/2}  
              \exp\left(t/\tau - r/\xi\right)
          & 1 \gg \alpha^\prime>0 \\
    \end{array} \right. 
\label{eq.Frt}
\end{equation}
where $\alpha^\prime=(\alpha-\alpha_c)/\alpha_c$ is the reduced control parameter, 
$\Gamma$ is the incomplete Gamma function and $\Psi$ is a scaling function 
defined by
\begin{equation}
\Psi(d,u)= \int_{u}^\infty\D z
    \frac{\Gamma\left(\frac{d}{2}-1,z\right)}{z^2}.
\end{equation}
Above the critical point the correlations diverge;
the time scale $\tau$ is given by Eq.~(\ref{eq.tau_mf})
and the correlation length by
\begin{equation}
\xi=\sqrt{\tau}=\sqrt{\frac{4A_2\alpha}{\alpha^\prime}}.
\end{equation}

Interestingly, as for $\alpha_c>0$ the correlations increase with time, what in usual
dynamical critical phenomena would be called the correlation 
time is {\em negative} while the correlation length is positive.
For $\alpha^\prime\le 0$ the dependence on $r^2/t$ directly shows that the
dynamical exponent is $z=2$. For $\alpha^\prime>0$ the time scale 
$\tau$ is the square of the length scale 
$\xi$, therefore also in this case
the dynamical exponent is $z=2$.

\begin{figure}[l]
\centerline{\epsfxsize=3.5in\epsfbox
{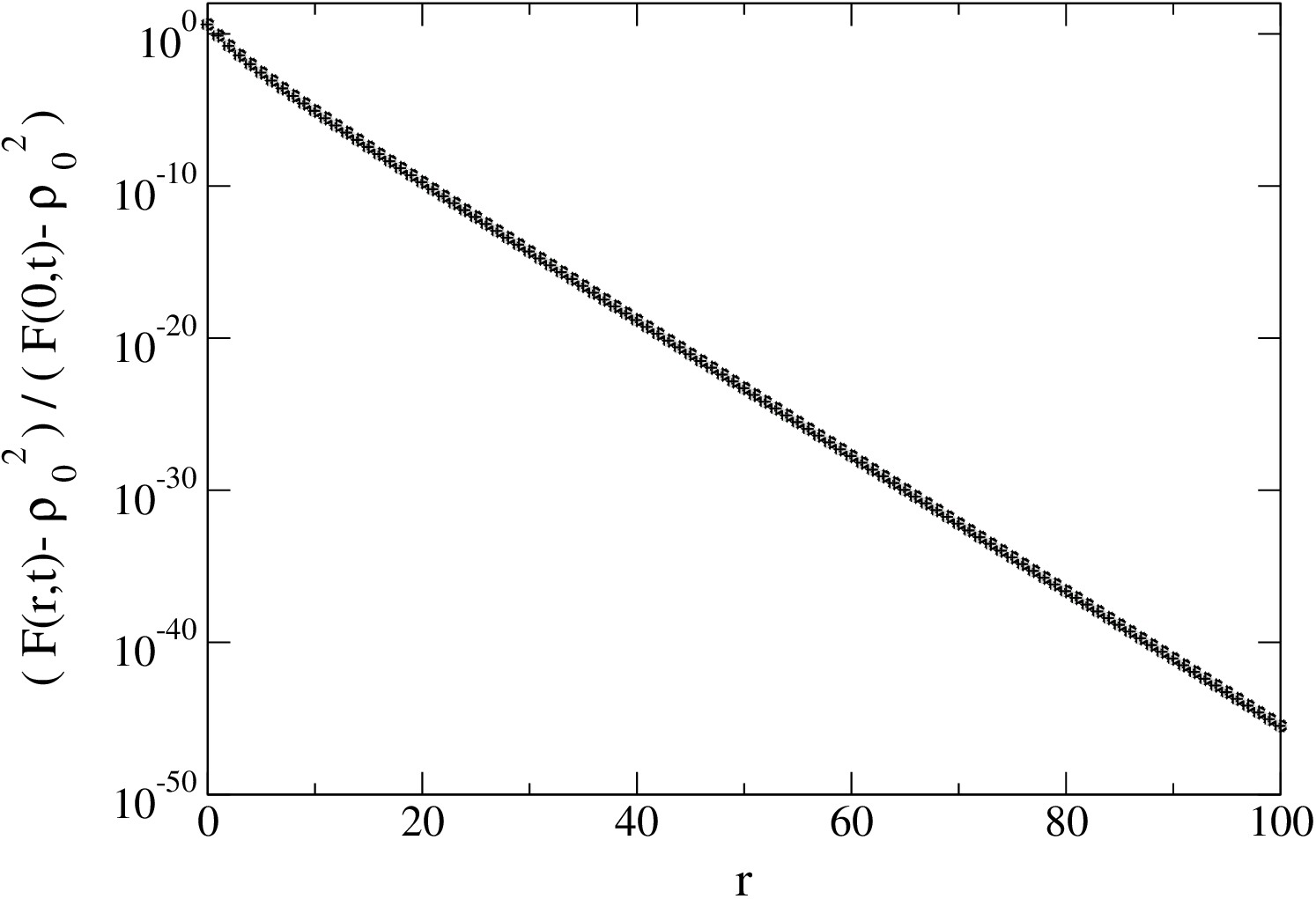}
}
\caption{Numerical calculation of $F({\bf r}=(r,0,...),t)$ for $d=3, \alpha=6>\alpha_c, \rho_0=1$ and
times $t=50,100,150,\ldots,400$.
\label{Fig4}}
\end{figure}

No analytical solution is available in the case $2<d<4$, thus we evaluate the
integral (\ref{eq.13}) numerically. 
Fig.~\ref{Fig3} shows the spatial dependence of the correlation function along
the axis ${\bf r}=(r,0,...)$, for $\alpha<\alpha_c$; Fig.~\ref{Fig4} shows
the case $\alpha>\alpha_c$.
A collapse of the calculated data points is achieved if we assume the following functional
dependence:  \\
\begin{equation}
F(r,t)-\rho_0^2\propto\left\{\begin{array}{cll}
      r^{2-d}f_1(r^2/t) & \alpha^\prime<0 \\[0.5cm]
      F(0,t) f_2(d,r) & \alpha^\prime>0,
    \end{array} \right.
\label{eq.27}
\end{equation}
where $f_1$ is a scaling function and $f_2(d,r)$ is a function that only depends on $d$ and $r$.
This result is in qualitative agreement with the previously derived formula for $d>4$.

%
%

\section{Discussion}

Apart from the spherical model this phase transition is related to a much 
simpler model: On a $d$--dimensional
cubic lattice non interacting particles are diffusing with rate $D$ and additionally
at site ${\bf x=0}$ particles may branch $A\to 2A$ with rate 
$\alpha^\prime=\alpha D$. 
The
equation for the time evolution of the particle density $\expect{n({\bf x},t)}$
is just given by Eq.~(\ref{eq.9}). We can adopt the solutions for $F({\bf r},t)$
by substituting the initial condition by $\rho_0^2 \rightarrow \rho_0$. In
particular we recover a phase transition for the particle
density at the origin. While in the original process it is rather complicated
to understand the physical meaning of the behavior of the second moment, in this
model we understand the behavior of the first moment: For $d=1$ diffusion
does not suffice to spread the particles on the lattice fast enough and the
particle density at ${\bf x=0}$ diverges for any given parameters. 
For higher dimensions additional spatial directions are accessible to spread
particles and as a consequence the particle density at ${\bf x=0}$ remains
finite for high enough diffusion constant $D$. 

The fact that the autocorrelation function is diverging while the particle
density remains constant allows some conclusions concerning the distribution
function for the particles $p(n)$ for late times. 
On the one hand, if $\expect{n}=\sum_{n} n\,p(n)$
is finite then for large $n$ the distribution function 
$p(n) < c_1\,n^{-\beta}$ with $\beta>2$. 
On the other hand, if $\expect{n^2}=\sum_{n} n^2\,p(n)$ 
is infinite then for large $n$ the distribution function
$p(n)> c_2 \,n^{-\beta}$ with $\beta<3$ with some positive constants $c_1, c_2$.
Thus the distribution function follows
for large $n$ a power law $p(n)\propto n^{-\beta}$ with $2<\beta<3$.

In summary, we have shown that for $d>2$ the bosonic PCPD exhibits a phase 
transition for $\expect{a({\bf x})^2}$ and thus for the autocorrelation function
$\sigma(t)^2=\expect{n({\bf x})^2}-\expect{n({\bf x})}^2=
\expect{a({\bf x})^2}+\expect{n({\bf x})}-\expect{n({\bf x})}^2$. The
order parameter $F_{\infty}^{-1}$ decreases linearly to zero for 
$\alpha\nearrow\alpha_c$ and is equal zero for $\alpha>\alpha_c$, where
$\alpha$ is proportional to the ratio of the reaction rates and the diffusion
constant. 
Thus diffusion has big influence in this process, it must be high enough in 
order to avoid a divergence of the autocorrelation. 

We have also shown that the critical properties of this process are related to the
mean spherical model.
As the spherical
model is a model for magnetism this analogy is rather intriguing
and the question arises whether it is just accidental.

\section*{Appendix}
\appendix*
For the meanfield case $d>4$ the solution of $F({\bf r},t)$ in the limit of large
${\bf r}$ and $t$ can derived analytically, as presented in what follows.

With the definition 
\begin{equation}
G({\bf r},t)=F({\bf r},t)-\rho_0^2
\end{equation}
the integral equation can be transformed to
\begin{equation}
G({\bf r},t)=\alpha\int_0^t \D\tau G({\bf 0},\tau) b({\bf r},t-\tau)+\alpha \rho_0^2
\int_0^t\D\tau b({\bf r},t).
\end{equation}
Using a Laplace transformation we get 
\begin{equation}
\tilde{G}({\bf r},p)=\alpha \tilde{G}({\bf{0}},p) \tilde{b}({\bf r},p)+\alpha\rho_0^2
\frac{\tilde{b}({\bf r},p)}{p},
\end{equation}
setting ${\bf r=0}$ determines $\tilde{G}({\bf 0},p)$ which yields
\begin{equation}
\tilde{G}({\bf r},p)=\alpha \rho_0^2 
\frac{\tilde{b}({\bf r},p)}{p\left(1-\alpha\tilde{b}({\bf 0},p)\right)}.
\end{equation}
The Fourier transform of this equation is
\begin{equation}
\tilde{g}({\bf q},p)=\rho_0^2\alpha \frac{1}{p\left(1-\alpha\tilde{b}({\bf 0},p)\right)}
\frac{1}{p+w({\bf q})}.
\end{equation}
For the meanfield case, $d>4$, $\tilde{b}({\bf 0},p)$ takes the simple form
\begin{equation}
\tilde{b}({\bf 0},p)=1/\alpha_c-p\gamma/\alpha,
\end{equation}
and we get
\begin{equation}
\tilde{g}({\bf q},p)=\frac{\rho_0^2\alpha}{\gamma}
        \frac{1}{p\left(p-\alpha^\prime/\gamma\right)}\frac{1}{p+w({\bf q})}.
\label{eq_gqp}
\end{equation}
Here we defined the reduced control parameter $\alpha^\prime=(\alpha-\alpha_c)/\alpha_c$
and  $\gamma=4 A_2\alpha$.
Although the function $\tilde{b}$ does not depend on dimension for $d>4$, generally a 
dependence of the solution $G({\bf r},t)$ on dimension is still possible as the inverse 
fourier transform depends on $d$, which does not affect the critical exponents.
Using 
an expansion into partial fractions we get
\begin{equation}
\tilde{g}({\bf q},p)=\rho_0^2\alpha\left(
  -\frac{1}{\alpha^\prime w({\bf q})}  \frac{1}{p}
  +\frac{1}{  \alpha^\prime\left(w({\bf q})
               +\alpha^\prime/\gamma\right)} \frac{1}{p-\alpha^\prime/\gamma}
   +\frac{1}{\gamma w({\bf q})\left(w({\bf q})+\alpha^\prime/\gamma\right)}\frac{1}{p+w({\bf q})}
\right).
\end{equation}
The inverse Laplace transform of this expression reads
\begin{equation}
g({\bf q},t)=\rho_0^2\alpha\left(-\frac{1}{\alpha^\prime w({\bf q})}
+\frac{1}{\alpha^\prime\left(w({\bf q})+\alpha^\prime/\gamma\right)} \exp\left(\frac{\alpha^\prime}{\gamma}t\right)
+\frac{1}{\gamma w({\bf q})\left(w({\bf q})+\alpha^\prime/\gamma\right)}
                           \exp\left(-w({\bf q}) t\right)
\right).
\label{eq_app8}
\end{equation}
Although the second term is not Laplace transformable for $\alpha^\prime>0$ this result
is correct and can be derived by transforming the function 
$H({\bf r},t)=\exp\left(-(\frac{\alpha^\prime}{\gamma}+\epsilon) t\right)G({\bf r},t)$
with $\epsilon>0$.

The inverse fourier transform of the first term of (\ref{eq_app8}) is
\begin{equation}
\begin{split}
\int\frac{\D^d {\bf q}}{(2\pi)^d}\, \e^{i{\bf qr}} \frac{1}{w({\bf q})}
&=\int_0^\infty \D x\int\frac{\D^d {\bf q}}{(2\pi)^d}\, \exp\left(-w({\bf q})x\right)\e^{i{\bf qr}}\\
&=\int_0^\infty \D x\, \e^{-2dx} I_{r_1}(2x)\cdot\ldots\cdot I_{r_d}(2x)\\
&\underset{\vert {\bf r} \vert \gg 1}{\approx} \int_0^\infty \D x\, (4\pi x)^{-d/2} \exp\left(-\frac{r^2}{4x}\right)\\
&=\frac{\Gamma\left(\frac{d}{2}-1\right)}{4\pi^{d/2}}r^{2-d}
\end{split}
\end{equation}

For the long time limit the second term contributes significantly only for $\alpha^\prime>0$. 
Defining $b^2=\alpha^\prime/\gamma$ we get for this case:
\begin{equation}
\begin{split}
\int\frac{\D^d {\bf q}}{(2\pi)^d}\frac{\e^{i{\bf qr}}}{w({\bf q})+b^2}
&=\int_0^\infty\D x \int\frac{\D^d {\bf q}}{(2\pi)^d} \exp\left(-\left(w({\bf q})+b^2\right)x\right) \e^{i{\bf qr}}\\
&=\int_0^\infty\D x \exp(-b^2 x)  \e^{-2dx} I_{r_1}(2x)\cdot\ldots\cdot I_{r_d}(2x)\\
&\underset{b^2 \ll 1,\vert {\bf r} \vert\gg 1}{\approx} (4\pi)^{-d/2} \int_0^\infty \D x\, x^{-d/2} \exp\left(-\frac{r^2}{4x}-b^2 x\right)\\
&=(4\pi)^{-d/2} \left(\frac{r^2}{4}\right)^{1-d/2} \int_0^\infty \D z\, z^{d/2-2}\exp\left(-z-\frac{b^2r^2}{4z}\right)\\
&=(4\pi)^{-d/2} \left(\frac{r^2}{4}\right)^{1-d/2} 2^{2-d/2}\left(br\right)^{d/2-1} K_{d/2-1}\left(br\right) \\
&\underset{r \gg 1}{\approx} (4\pi)^{-d/2} \left(\frac{r^2}{4}\right)^{1-d/2} 2^{2-d/2}\left(br\right)^{d/2-1} \sqrt{\frac{\pi}{2}}
\frac{\exp(-br)}{\sqrt{br}} \\
&=\frac{1}{2^{(d+1)/2}\pi^{(d-1)/2}} b^{(d-3)/2} r^{(1-d)/2} \exp(-br)\\ 
&=\frac{1}{2^{(d+1)/2}\pi^{(d-1)/2}} \left(\frac{\alpha^\prime}{\gamma}\right)^{(d-3)/4} r^{(1-d)/2} 
\exp\left(-\sqrt{\frac{\alpha^\prime}{\gamma}}r\right),
\end{split}
\end{equation} 
where $K_{d/2-1}$ is the modified Bessel function of second kind.

For $\alpha^\prime>0$ the third term is transformed to:
\begin{equation}
\begin{split}
\int\frac{\D^d {\bf q}}{(2\pi)^d}&\frac{\e^{i{\bf qr}}\exp\left(-w({\bf q})t\right)}{w({\bf q})\left(w({\bf q})+b^2\right)}\\
&= \int_0^\infty\D x \int\frac{\D^d {\bf q}}{(2\pi)^d} \exp\left(-(w({\bf q})+b^2)x\right)
    \frac{ \e^{i {\bf qr}} \exp\left(- w({\bf q}) t\right)}{w({\bf q})} \\
&=  \int_0^\infty\D x \exp(-b^2 x) \int_{t+x}^\infty \D y 
    \int\frac{\D^d {\bf q}}{(2\pi)^d} \exp\left(-w({\bf q}) y\right) \e^{i{\bf qr}}\\
&=  \int_0^\infty\D x \exp(-b^2 x) \int_{t+x}^\infty \D y  
    \,\e^{-2dy} I_{r_1}(2y)\cdot\ldots\cdot I_{r_d}(2y) \\
&\underset{\vert {\bf r} \vert \gg 1}\approx  (4\pi)^{-d/2} \int_0^\infty\D x \exp(-b^2 x) \int_{t+x}^\infty \D y 
   \,y^{-d/2} \exp\left(-\frac{r^2}{4y}\right)\\
&=   (4\pi)^{-d/2} t^{1-d/2} \left(\frac{r^2}{4t}\right)^{1-d/2} \int_0^\infty\D x \exp(-b^2 x)
    \int_0^{\frac{r^2}{4t(1+x/t)}} \D z\,z^{d/2-1}\exp(-z) \\
&\underset{t \gg 1}{\approx} \frac{r^{2-d} }{4\pi^{d/2}}   \int_0^\infty\D x \exp(-b^2 x)
     \left(  \int_0^{\frac{r^2}{4t}}\D z\, z^{d/2-1}\exp(-z)-\left(\frac{r^2}{4t}\right)^{d/2-1}
            \exp(-\frac{r^2}{4t}) \frac{x}{t} \right) \\
&= \frac{r^{2-d} }{4\pi^{d/2}}   \int_0^\infty\D x \exp(-b^2 x)
     \left( \Gamma\left(\frac{d}{2}-1\right)-\Gamma\left(\frac{d}{2}-1,\frac{r^2}{4t}\right) -\left(\frac{r^2}{4t}\right)^{d/2-1}
            \exp(-\frac{r^2}{4t}) \frac{x}{t} \right) \\
&= \frac{r^{2-d} }{4\pi^{d/2}}b^{-2}\left( \Gamma\left(\frac{d}{2}-1\right)-
         \Gamma\left(\frac{d}{2}-1,\frac{r^2}{4t}\right) -
        \frac{b^{-2}}{t}\left(\frac{r^2}{4t}\right)^{d/2-1} \exp(-\frac{r^2}{4t})  \right),
\end{split}
\end{equation}
with the incomplete Gamma-function $\Gamma(a,x)=\int_x^\infty \D t \, t^{a-1} \e^{-t}$.

For $\alpha^\prime<0$ the same result can be derived by substituting $x\to -x$.

At the critical rate $\alpha^\prime=0$ equation (\ref{eq_gqp}) reduces to
\begin{equation}
\tilde{g}({\bf q},p)=\frac{\rho_0^2\alpha}{\gamma}
        \frac{1}{p^2}\frac{1}{p+w({\bf q})}.
\end{equation}
The inverse Laplace transform of this expression is given by
\begin{equation}
\begin{split}
{\cal L}^{-1}\left(\frac{1}{p^2\left(p+w({\bf q})\right)},t\right)&=
    \int_0^t \D\tau\,\int_0^\tau\D\tau^\prime 
        {\cal L}^{-1}\left(\frac{1}{p+w({\bf q})},\tau^\prime\right)\\
&= \int_0^t \D\tau\int_0^\tau\D\tau^\prime\, \exp(-w({\bf q})\tau^\prime).
\end{split}
\end{equation}
The necessary conditions for this equality are fulfilled \cite{grad}:
\begin{equation}
\begin{split}
\lim_{t\to\infty}\left(\e^{-pt}\int_0^t\D\tau \exp(-w({\bf q})\tau)\right)&=0 \\
\lim_{t\to\infty}\left(\e^{-pt}\int_0^t\D\tau 
                              \int_0^\tau\D\tau^\prime\exp(-w({\bf q})\tau^\prime)\right)&=0. 
\end{split}
\end{equation}
This yields
\begin{equation}
\begin{split}
G({\bf r},t)&=\frac{\rho_0^2\alpha}{\gamma} \int_0^t \D\tau\int_0^\tau\D\tau^\prime\, 
         \int\frac{\D^d {\bf q}}{(2\pi)^d} \exp(-w({\bf q})\tau^\prime)\\
&= \frac{\rho_0^2\alpha}{\gamma} \int_0^t \D\tau\int_0^\tau\D\tau^\prime\,  
         \e^{-2dx} I_{r_1}(2x)\cdot\ldots\cdot I_{r_d}(2x) \\
&\underset{\vert {\bf r} \vert \gg 1}{\approx} \frac{\rho_0^2\alpha}{\gamma} \int_0^t \D\tau\int_0^\tau\D\tau^\prime\, 
   (4\pi\tau^\prime)^{-d/2} \exp\left(-\frac{r^2}{4\tau^\prime}\right) \\
&=  \frac{\rho_0^2\alpha}{\gamma(4\pi)^{d/2}} \left(\frac{r^2}{4}\right)^{-d/2+1} \int_0^t\D\tau
    \int_{\frac{r^2}{4\tau}}^{\infty} \D z z^{d/2-2} \exp(-z) \\
&=  \frac{\rho_0^2\alpha}{4\gamma \pi^{d/2}} r^{2-d} \int_0^t\D\tau
    \Gamma\left(\frac{d}{2}-1,\frac{r^2}{4\tau}\right)\\
&=\frac{\rho_0^2\alpha}{16\gamma \pi^{d/2}}\,\,r^{4-d}\,\, \int_{\frac{r^2}{4t}}^\infty\D z
    \frac{\Gamma\left(\frac{d}{2}-1,z\right)}{z^2}
\end{split}
\end{equation}

Thus we get the expressions Eq.~(\ref{eq.Frt}) in the limit of large ${\bf r}$ and $t$.

\section*{Acknowledgments}

We thank the Max--Planck Institut f\"ur komplexe Systeme (MPIPKS) in Dresden
for warm hospitality, where parts of this work were done.

\end{document}